\documentclass[%
 reprint,
 amsmath,amssymb,
 aps,
]{revtex4-1}

\usepackage{dcolumn}% Align table columns on decimal point
\usepackage{bm}% bold math
\usepackage[mathlines]{lineno}

\newtheorem{theorem}{Theorem}
\newtheorem{corollary}{Corollary}[theorem]
\newtheorem{lemma}{Lemma}

\begin{document}

\preprint{APS/123-QED}

\title{Probabilistic Quantum Teleportation}
\author{Jianhao M. Yang}
%\altaffiliation[Also at ]{Qualcomm Inc.}%Lines break automatically or can be forced with \\
\email{jianhao.yang@alumni.utoronto.ca}
\affiliation{
Qualcomm, San Diego, CA 92121, USA
}

\date{\today}

\begin{abstract}
Teleporation with partially entangled quantum channel cannot achieve unit fidelity and unit probability. We show that the condition for faithful teleportation of a pure state or a mixed state can be described by $\rho_q\rho_m =pI$, where $\rho_q$ is the reduced density matrix of the quantum channel, $\rho_m$ is the reduced density matrix of the measurement basis, and $p$ is the probability of faithful teleportaion. We investigate the invariance of faithful teleportation conditions under unitary transformation. These results not only bring new insights to the probabilistic quantum teleportation theory,  but also offer operational significance in that a simple procedure is provided to find out the faithful teleportaiton probability and the matching measurement basis for any partially entangled quantum channel.
\begin{description}
\item[PACS numbers] 03.65.Ud, 03.67.-a
\end{description}
\end{abstract}
\pacs{03.65.Ud, 03.67.-a}
\maketitle

\section{\label{intro}Introduction}

Entanglement is one of the most distinct features of quantum mechanics. One of the applications of entanglement in quantum information technology is quantum teleportaion~\cite{Bennett93}. When two particles A and B are entangled, none of them has a definite quantum state. Instead each of the particles is in mixed state and be described with a reduced density matrix. The quantum teleporation protocol explores this indeterminacy as a communication resource. Through a Bell state measurement performed by Alice on particle A and a third particle C, and the classical communication from Alice to Bob, the mixed state of particle B can be transformed into be a pure state that replicates the unknown quantum state of particle C. The two classical bits that Alice sent to Bob enable Bob to complete this transformation. In this protocol, the entanglement between particles A and B is the prerequisite for the success of teleportation. Quantum teleporation is an important element for quantum communication and quantum computing, such as quantum repeater~\cite{Briegel98}, quantum gate teleportatin~\cite{Gottesman99}, quantum network~\cite{Raussendorf01}, and measurement-base computing. Ref~\cite{Pirandola15} provides a comprehensive survey of the most recent advancement of quantum teleportation. 

In the original teleportation protocol, the shared particles between Alice and Bob are in maximum entangled state, and the Bell states which Alice performs the measurement are also maximum entangled. In this scheme, the teleportion achieves unit fidelity and unit probability. The protocol has been extended to many variants, including entanglement swapping~\cite{Bennett93}, teleporting states with more than two dimensions~\cite{Werner01} or state with continuous variables~\cite{Braunstein98}, and teleportation involved more than two qubits~\cite{Karlsson98}. To ensure Bell states are distinguishable, different methods were introduced such as complete hyperentangled Bell state analysis~\cite{Kwiat98, Sheng12}. However, in reality, maximum entangled state is very difficult to maintain as it evolves and interacts with the environment. Perfectly entangled state is also difficult to generate. For practical purpose it is important to study teleportation using partially entangled quantum channel. Using perfectly entangled particle should be considered as a special case. 

There are already extensive researches on the teleportation scheme when the shared resource between Alice and Bob are in partially entangled states. For example, the partially entangled particles can be concentrated or purified into maximum entangled states at the cost of reducing the total number of partially entangled particles~\cite{Bennett96}. One can also use higher dimensional entangled particles~\cite{Gour04}, or add an auxiliary qubit~\cite{Kurucz03}, etc. The common theme of these approaches is to introduce cost of additional resources. Probabilistic teleportation, on the other hand, does not require additional resource. Instead, it just relies on the given partially entangled resource but accepts the fact that faithful teleportaion is only successful with less than unit probability. Theoretical description of the probabilistic teleporation using generic von-Neumann measurement instead Bell state measurement can be found in~\cite{Li00, Kurucz01, Agrawal02, Albeverio03}. 

A basic question is that when the quantum channel is in partially entangled state, how the faithful teleporation probability is quantified? Intuitively this probability should be related to the degree of entanglement of the quantum channel, and related to the degree of entanglement in the chosen measurement basis. It is desirable to formulate the relation among them. Furthermore, in most of formulations of teleportation theory, it is typically assumed that Bob will perform a special unitary operation to recover the state of particle C, such as the unit matrix or the Pauli matrix. This assumption needs to be relaxed in probabilistic teleportation as long as the unitary operator is independent of the unknown state of particle C.

The purpose of this article is to formulate the general conditions for faithful teleportation and derive the  probability of faithful teleportation. An elegant relation among the reduced density matrix of the quantum channel, the reduced density matrix of measurement basis, and the faithful teleportation probability is presented. We further confirm the invariance of the teleportation conditions and probability under unitary transformation. These results not only address the questions mentioned earlier, but also bring operational significance which offers a simple procedure to find out the faithful teleportaiton probability and the matching measurement basis for any partially entangled quantum channel.

The paper is organized as following. In section \ref{generalconditions}, the conditions for achieving faithful teleportation are formulated. Section \ref{unitaryT} and \ref{mixedstate} show that such conditions are invariant under unitary transformation, and they are the same regardless the teleported quantum state is a pure state or a mixed state. In Section \ref{NEqual2}, we provide a rigorous answer to the question that for any set of orthogonal measurement basis, how many of them lead to faithful teleportation assuming Bob can perform any form of unitary operation. We then apply the theory to concrete examples and discuss the physical and operational implications.

\section{\label{generalconditions}Conditions for faithful teleportation}
Assume Alice and Bob share a particle pair A and B, where Alice has particle A and Bob has particle B. The pair is described by a pure state in ${\cal H}_A\otimes {\cal H}_B$, where $\dim {\cal H}_A=\dim {\cal H}_B=N$. Let $\{|j\rangle_A\}$ and
$\{|i\rangle_B\}\ (i, j=0,\ldots ,N-1)$ the orthogonal bases on ${\cal
  H}_A$ and ${\cal H}_B$, respectively. The state of the shared pair is
\begin{equation}
  \label{partent}
|\Psi\rangle_{AB}=\sum_{i,j=0}^{N}q_{ij}|i\rangle_A|j\rangle_B, \quad \sum_{ij}|q_{ij}|^2=1
\end{equation}
Alice also has a qubit C in pure state  $|\Psi\rangle_C=\sum_{k=0}^{N-1}c_k|k\rangle_C$. The combined three particle system then can be described as
\begin{equation}
  \label{3partie}
|\Psi\rangle_{ABC}=\sum_{i,j,k=0}^{N-1}q_{ij}c_k|i\rangle_A|j\rangle_B|k\rangle_C
=\sum_{i,j,k=0}^{N-1}q_{ij}c_k|ik\rangle_{AC}|j\rangle_B
\end{equation}
Now Alice performs a general von-Neumann measurement on the two particles AC in her lab. In order to be able to distinguish each measurement, the measurement basis must be orthogonal. For the pair AC, there can be $N^2$ orthogonal measurement basis. Each of such measurement basis, denoted as $|\psi^m_{AC}\rangle$ where $(m=0, \ldots, N^2)$, is given by
\begin{equation}
  \label{mbasis}
|\psi^m_{AC}\rangle=\sum_{i,k=0}^{N-1}d_{m,ik}|ik\rangle_{AC}
\end{equation}
and the parameter $d_{m,ik}$ forms a unitary matrix with dimension of $N^2$, i.e.
\begin{equation}
  \label{orthogonal}
\langle\psi^n_{AC}|\psi^m_{AC}\rangle=\sum_{i,k=0}^{N-1}d^*_{n,ik}d_{m,ik}=\delta_{mn}
\end{equation}
Inverting Eq.(\ref{mbasis}), we have $|ik\rangle_{AC}=\sum_md^*_{m,ik}|\psi^m_{AC}\rangle$, substitute this into Eq.(\ref{3partie}), 
\begin{equation}
  \label{3partie2}
|\Psi\rangle_{ABC}=\sum_m(\sum_{i,j,k=0}^{N-1}q_{ij}c_kd^*_{m,ik}|j\rangle_B)|\psi^m_{AC}\rangle
\end{equation}
When Alice performs von-Neumann measurement with the basic $|\psi^m_{AC}\rangle$, the Schmidt projection of the wavefunction is
\begin{equation}
  \label{particleB}
  \begin{split}
|\Psi\rangle^m_{ABC} 
& = p_m^{-1/2}|\psi^m_{AC}\rangle \langle\psi^m_{AC}|\Psi\rangle_{ABC} \\
& =p_m^{-1/2}|\psi^m_{AC}\rangle\sum_{i,j,k=0}^{N-1}q_{ij}c_kd^*_{m,ik}|j\rangle_B
\end{split}
\end{equation}
where $p_m=\left \|\sum_{i,j,k}q_{ij}c_kd^*_{m,ik}|j\rangle_B\right \|^2$ is the probability of measurement outcome. Let's rewrite $d_{m,ik}=d^m_{i,k}$, and define $D_m$ as an $N$ dimensional matrix with element $d^m_{i,k}$, note that $m$ is just a label here, and ${i,k}$ are the index for the matrix elements. Matrix $D_m$ describes the measurement and the outcome will be communicated to Bob via classical channel, therefore it describes the LOCC. We also define $Q$ as an $N$ dimensional matrix with and element $q_{ij}$. Essentially, matrix $Q$ describes the quantum channel. We further define a product matrix $L_m=Q^TD_m^*$ where $Q^T$ is the transpose of $Q$, then $(L_m)_{jk}=\sum_iq_{ij}d^*_{m,ik}$. With these notations, the state o particle B in Eq.(\ref{particleB}) becomes (ignoring the state of Alice's particles A and C)
\begin{equation}
  \label{particleB2}
  \begin{split}
|\Psi\rangle_B
& =p_m^{-1/2}\sum_{j,k=0}^{N-1}(L_m)_{jk}c_k|j\rangle_B\\
& =p_m^{-1/2}\sum_j(\sum_k(L_m)_{jk}c_k)|j\rangle \\
& =p_m^{-1/2}L_m|\Psi\rangle_C = p_m^{-1/2}Q^TD^*_m|\Psi\rangle_C
\end{split}
\end{equation}
This is the wave function for Bob's particle. For successful teleportation, Bob can just perform a unitary operation $U_m$ to recover the original quantum state of particle C, i.e., $U_m|\Psi\rangle_B=p_m^{-1/2}U_mQ^TD^*_m|\Psi\rangle_C=|\Psi\rangle_C$. Therefore $p_m^{-1/2}U_mQ^TD^*_m = I$. This gives the condition for successful teleportation for the von-Neumann measurement with basis $|\psi^m_{AC}\rangle$, assuming matrix $Q$ is invertible,
\begin{equation}
\label{condition1}
D_m = p_m^{1/2}(Q^{-1})^\dag U^T_m
\end{equation}
Using the notation of $D_m$, we can rewrite Eq.(\ref{orthogonal}) as
\begin{equation}
\label{condition2}
Tr(D_n^{\dag}D_m) = \delta_{mn}
\end{equation}
substitute $D_m$ in Eq.(\ref{condition1}) into Eq.(\ref{condition2}) , we get 
\begin{equation}
\label{condition22}
\begin{split}
Tr(D_n^{\dag}D_m) & =\sqrt{p_mp_n}Tr(Q^{-1}(Q^{-1})^\dag U_m^TU_n^*) \\
& =\sqrt{p_mp_n}Tr((Q^\dag Q)^{-1}U_m^TU_n^*) = \delta_{nm}
\end{split}
\end{equation}
Denote matrix $M=Q^{\dag}Q$ which is a Hermitian matrix.  Let $n=m$ and note that $U_m^TU_m^*=I^*=I$, we obtain the probability when faithful teleportation is successful, 
\begin{equation}
\label{probability}
p=(Tr(M^{-1}))^{-1}
\end{equation}
Note that $p$ is independent of $m$. Eq.(\ref{condition2}) and (\ref{probability}) are similar to the results in Ref.~\cite{Kurucz01} for faithful teleportaion condition. However, not every $D_m$ satisfies both Eq. (\ref{condition1}) and (\ref{condition2}). Let's define that for any given orthogonal measurement basis $|\psi^m_{AC}\rangle$ where $(m=0, \ldots, N^2)$, the number of measurements that satisfied faithful teleportation condition Eq. (\ref{condition1}) is $\eta$. 
There are unlimited number of von-Neumann measurements that can satisfy the faithful teleportation condition (\ref{condition1}). However, when a set of orthogonal von-Neumann measurements is chosen, only $\eta$ of them can satisfy faithful teleportation condition. Furthermore, since the state of particle C is unknown to Alice and Bob, the choice of $U_m$ and therefore $U_m$ should be independent of $|\Psi\rangle_C$. The question we want to address here is what maximum $\eta$ can be found when using partially entangled resource as the quantum channel. 

\section{\label{unitaryT}Invariance under unitary transformation}
In this section we show that teleportation conditions and the probability are invariance under unitary transformation. 

Initially the wave function for particles A, B, or C are described by the orthogonal basis $\{|i\rangle_A\}$, $\{|j\rangle_B\}$, and  $\{|k\rangle_C\} (i,j,k=0,N)$ in picture ${\cal F}$. According to the Dirac transformation theory, we can re-express the state with a new orthogonal basis  $\{|l\rangle_A\}$, $\{|m\rangle_B\}$, and  $\{|n\rangle_C\} (l, m, n=0,N)$ in picture ${\cal G}$, where ${\cal F}$ and ${\cal G}$ are related through unitary transformation. Denote unitary matrix $(U_A)_{li}=\langle l|i\rangle_A$, $(U_B)_{mj}=\langle m|j\rangle_B$, and $(U_C)_{nk}=\langle n|k\rangle_C$, we can re-express Eq.(\ref{partent}) as
\begin{equation}
  \label{entgeneralG}
  \begin{split}
|\Psi\rangle_{AB} 
& = \sum_{i,j}\sum_{l,m}q_{ij}(U_A)_{li}(U_B)_{mj}|l\rangle_A|m\rangle_B \\
& =\sum_{l,m}(\sum_j(\sum_i(U_A)_{li}q_{ij})(U_B)_{mj})|l\rangle_A|m\rangle_B
\end{split}
\end{equation}
Recall in section II we have defined matrix $Q$ in ${\cal F}$ with element $q_{ij}$, here we can also introduce matrix $Q^\prime$ in ${\cal G}$ with element $q^\prime_{lm}=\sum_j(\sum_i(U_A)_{li}q_{ij})(U_B)_{mj}$ such that $|\Psi\rangle_{AB}=\sum_{l,m}q^\prime_{lm}|l\rangle_A|m\rangle_B$. This means $Q^\prime=U_AQU_B^T$ where $U_B^T$ is the transpose of $U_B$. Based on the well-known singular value decomposition, any $N\times N$ square complex matrix $Q$ can be diagonalized by two unitary matrices, i.e., $U_LQU_R^\dag=Q^\prime$ where $Q^\prime$ is diagonal. If we choose $U_A=U_L$, and $U_B=U_R^*$, so that $U_B^T=U_R^\dag$ (note that if $U_B$ is unitary, $U_B^T$ is also unitary), we find a transformation ${\cal F\to \cal G}$ such that $Q^\prime$ is diagonal. Therefore, for entangled state described in picture ${\cal F}$ by Eq. (\ref{partent}), there exists a transformation ${\cal F\to \cal G}$ such that the same state is expressed as
\begin{equation}
  \label{entG}
|\Psi\rangle^{\cal G}_{AB}=\sum_{i=0}^{N}q^\prime_{i}|i\rangle_B|i\rangle_A, \quad \sum_{i}|q^\prime_{i}|^2=1
\end{equation}

Similarly, for matrix $D_m$ that describes the orthogonal measurement basis Alice performs on particle A and C, after ${\cal F\to \cal G}$, the matrix becomes $D^\prime_m=U_AD_mU_C^T$.
\begin{lemma}
The teleportaiton conditions and the probability of faithful teleportation are invariants under unitary transformation ${\cal F\to \cal G}$.
\end{lemma}
It is trivial to show the orthogonal condition is preserved. Substitute $D_m^\prime=U_AD_mU_C^T$ into Eq.(\ref{condition2}), 
\begin{equation*}
  \label{orthogonalG}
  \begin{split}
  Tr((D^\prime_n)^{\dag}D^\prime_m) &=Tr((U_AD_nU_C^T)^\dag(U_AD_mU_C^T))\\
  &=Tr((U_C^T)^\dag(U_A^\dag U_A)(D_n^\dag D_m)U_C^T)\\
  &=Tr(D_n^\dag D_m) = \delta{mn}
\end{split}
\end{equation*}
For the teleportation condition, Eq.(\ref{particleB2}) after transformation ${\cal F\to \cal G}$ becomes
\begin{equation}
  \label{particleBG}
  \begin{split}
|\Psi\rangle_B^m
& = (p^\prime_m)^{-1/2}(Q^\prime)^T (D_m^\prime)^*|\Psi\rangle_C^{\cal G} \\
& = (p^\prime_m)^{-1/2}U_BQ^TU_A^TU_A^*D_m^*U^{\dag}_CU_C|\Psi\rangle_C^{\cal F}\\
& = (p^\prime_m)^{-1/2}U_BQ^TD^*_m|\Psi\rangle_C^{\cal F} \\ 
& = \sqrt{(p_m/p^\prime_m)}U_BU_m^{\dag}|\Psi\rangle_C^{\cal F}
\end{split}
\end{equation}
Bob needs to perform unitary operation $U_mU_B^\dag$. Since $\langle\Psi|\Psi\rangle_B^m=1$, it is obvious that $p_m^\prime=p_m$.
This proof can be shown in another way. When Eq.(\ref{condition1}) is true, Bob's particle after receiving the classical information is described by Eq.(\ref{particleB2}), i.e., $|\Psi\rangle_B^m=\langle\psi^m_{AC}|\Psi_{ABC}\rangle^{\cal F} =U_m^\dag |\Psi\rangle_C^{\cal F}$. After transformation ${\cal F\to \cal G}$,
\begin{equation*}
  \label{eq:orthogonal}
  \begin{split}
  |\Psi\rangle_B^m & =\langle\psi^m_{AC}|\Psi_{ABC}\rangle^{\cal G}\\
&= \langle\psi^m_{AC}|(U_A\otimes U_C)^\dag (U_A\otimes U_B\otimes U_C)|\Psi_{ABC}\rangle^{\cal F} \\
&= U_B\langle\psi^m_{AC}|\Psi_{ABC}\rangle^{\cal F} =U_BU_m^\dag|\Psi\rangle_C^{\cal F}
\end{split}
\end{equation*}
Again, Bob needs to perform unitary operation $U_mU_B^\dag$ to recover the quantum state of particle C. However, Eq.(\ref{particleBG}) has the advantage of showing $p_m^\prime=p_m$. 

As will be seen in Section \ref{NEqual2}, the number of orthogonal measurement basis that faithfully teleport a qubit, $\eta$, strongly depends on the whether the shared pair is in maximum entangled state or partially entangled state. It is important to confirm that the degree of entanglement is unchanged under the unitary transformation. 
\begin{lemma}
The degree of entanglement for the entangled pair is an invariant under unitary transformation ${\cal F\to \cal G}$.
\end{lemma}
The degree of entanglement for the entangled pair A and B can be measured by the von Neumann entropy of the reduced density matrix of A or B. For entangled pair described by Eq. (\ref{partent}), denote the reduced density matrix for A as $\rho^{\cal F}_A$, its matrix element is $(\rho^{\cal F}_A)_{ij}=\sum_kq_{ik}q_{jk}^* = (QQ^{\dag})_{ij}$. Therefore $\rho^{\cal F}_A=QQ^{\dag}$. After ${\cal F\to \cal G}$, since $Q^\prime=U_AQU_B^T$, we have $\rho_A^{\cal G}=Q^\prime(Q^\prime)^\dag =U_AQU_B^TU_B^*Q^\dag U_A^\dag = U_A\rho^{\cal F}_AU_A^\dag$. This means $\rho_A^{\cal G}$ and $\rho_A^{\cal F}$ are similar matrices and have the same eigenvalue sets $\{\lambda_i\} (i=0,\ldots, N-1)$. The von-Neumann entropy $E=-\sum_i\lambda_iln\lambda_i$ is therefore an invariant.

Given an orthogonal measurement basis set, in order to find the number of measurement basis that satisfy the faithful teleportation conditions, $\eta$, we just need to find the number of unitary matrices that satisfy Eq. (\ref{condition1}) and Eq. (\ref{condition2}). The good news is that $\eta$ is also an invariance under unitary transformation.
\begin{theorem} 
\label{theoryeta}
$\eta$ is an invariant when an arbitrary entangled state is transformed to its Schmidt decomposition form (\ref{entG}) using unitary operation.
\end{theorem}
To prove, let's assume among the orthogonal set $|\psi^m_{AC}\rangle$ where $(m=0, \ldots, N^2)$, $\eta$ of them satisfy the faithful teleportaiton condition Eq.(\ref{condition1}). The entangled quantum channel, described by Eq. (\ref{partent}), is then transformed to the Schmidt decomposition form (\ref{entG}) through unitary matrix. Based on Lemma 1, in the new picture ${\cal G}$, the orthogonal set $|\psi^m_{AC}\rangle^\prime$ are still orthogonal each other, and any measurement that satisfies Eq. (\ref{condition1}) in picture ${\cal F}$ still satisfies the faithful teleportation condition. Therefore we find at least $\eta$ orthogonal measurement basis that satisfies Eq. (\ref{condition1}), i.e., after transformation, the total number of orthogonal measurement basis that satisfies Eq. (\ref{condition1}) $\eta^\prime \ge \eta$. Now we transform the Schmidt decomposition form (\ref{entG}) back to Eq. (\ref{partent}). Based on the same Lemma 1, we have $\eta \ge \eta^\prime$. Therefore $\eta^\prime = \eta$. Lemma 2 shows that after the transformation, a partially entangled shared pair AB  is still in partially entangled states, and a maximum entangled shared pair AB is still in maximum entangled states. This ensures $\eta$ is not changed due to the change of degree of entangelemt.    

\section{\label{mixedstate} Density Matrix Formulation and the main results}
If Alice wants to teleport a mixed state instead of a pure state, do the faithful teleportation conditions remains the same? Intuitively this should be the case since the faithful teleportation conditions are independent of the state Alice wants to teleport. We will show this is indeed true in this section. Reformulating the teleportaiton condition using density matrix gives more generic results, and as will be shown, it uncovers the relation among the reduced density matrix of the quantum channel, the reduced density matrix of measurement basis, and the faithful teleportation probability in a very simple form.

Assume the mixed state Alice wants to teleport is described by
\begin{equation}
\label{mixedstateC}
\hat{\rho}_C=\sum_{k=0}^{N}\rho_{k}|k\rangle_C\langle k| \quad (\sum_{k}\rho_{k} =1, 0<\rho_{k}<1)
\end{equation}
We can rewrite Eq.(\ref{mixedstateC}) in a more general form
\begin{equation}
\label{mixedstateC2}
\hat{\rho}_C=\sum_{k,k^\prime=0}\rho_{kk^\prime}|k\rangle_C\langle k^\prime|
\end{equation}
the corresponding density matrix is $\rho_C$ with element $\rho_{kk^\prime}$. When $\rho_C^2 \ne \rho_C$ it is a mixed state. The combined system of the shared entangled pair and the mixed state to be teleported is
\begin{equation}
\label{mixedsystem}
\begin{split}
&\hat{\rho}_{ABC} =|\Psi\rangle_{AB}\langle\Psi|\otimes\hat{\rho}_C \\
& =\sum_{ij=0}^{N}\sum_{i^\prime j^\prime=0}^{N}q_{ij}q^*_{i^\prime j^\prime}|ij\rangle_{AB}\langle i^\prime j^\prime| \otimes\sum_{kk^\prime=0}^{N}\rho_{kk^\prime}|k\rangle\langle k^\prime|\\
&=\sum_{ij}\sum_{i^\prime j^\prime}\sum_{kk^\prime}q_{ij}q^*_{i^\prime j^\prime}\rho_{kk^\prime}(|ik\rangle_{AC}\langle i^\prime k^\prime|)\otimes |j\rangle_B\langle j^\prime|
\end{split}
\end{equation}
Now Alice performs a general von-Neumann measurement, as defined by the orthogonal basis Eq.(\ref{mbasis}),  on the two particles AC in her lab. With these orthogonal basis,
\begin{equation*}
|ik\rangle_{AC}\langle i^\prime k^\prime|=\sum_{mm^\prime=0}^{N^2}d^{m*}_{ik}d^{m^\prime}_{i^\prime k^\prime}|\Psi_{AC}^m\rangle\langle \Psi_{AC}^{m^\prime}|
\end{equation*}
Substitute this into Eq.(\ref{mixedsystem}), we get
\begin{equation}
\label{mixedsystem2}
\begin{split}
\hat{\rho}_{ABC} = &\sum_{mm^\prime}(\sum_{ij}\sum_{i^\prime j^\prime}\sum_{kk^\prime}q_{ij}q^*_{i^\prime j^\prime}\rho_{kk^\prime}d^{m*}_{ik}d^{m^\prime}_{i^\prime k^\prime}\\
&|j\rangle_B\langle j^\prime|) \otimes |\Psi_{AC}^m\rangle\langle \Psi_{AC}^{m^\prime}|
\end{split}
\end{equation}
After Alice performs the von-Neumann measurement with eigenstate $\Psi_{AC}^m$, and Bob receives the result through LOCC, the state of particle B is transformed into
\begin{equation}
\label{mixedstateB}
\begin{split}
\hat{\varrho}_B & = \langle I\otimes\Psi_{AC}^m| \hat{\rho}_{ABC} |I\otimes\Psi_{AC}^m\rangle \\
& = p_m^{-1}\sum_{ij}\sum_{i^\prime j^\prime}\sum_{kk^\prime}q_{ij}q^*_{i^\prime j^\prime}\rho_{kk^\prime}d^{m*}_{ik}d^{m^\prime}_{i^\prime k^\prime}|j\rangle_B\langle j^\prime|
\end{split}
\end{equation}
where $p_m$ is the probability of observing the measurement outcome and determined by $Tr(\hat{\varrho}_B)=1$. In section \ref{generalconditions}, we have defined a product matrix $L=Q^TD^*$ such that $(L_m)_{jk}=\sum_iq_{ij}d^*_{m,ik}$. This means $L^\dag=D^TQ^*$ with $(L_m^{\dag})_{k^\prime j^\prime}=\sum_{i^\prime}q^*_{i^\prime j^\prime}d^{m^\prime}_{i^\prime k^\prime}$. With these notations, the summations over $ii^\prime$ and $kk^\prime$ in Eq.(\ref{mixedstateB}) are simplified as
\begin{equation}
\begin{split}
\hat{\varrho}_B & = p_m^{-1}\sum_{jj^\prime}\sum_{kk^\prime}(L_m)_{jk}\rho_{kk^\prime}(L_m^{\dag})_{k^\prime j^\prime}|j\rangle_B\langle j^\prime| \\
& = p_m^{-1}\sum_{jj^\prime} (L_m\rho_C L_m^\dag)_{jj^\prime} |j\rangle_B\langle j^\prime|
\end{split}
\end{equation}
The density matrix $\varrho_B = L_m\rho_C L_m^\dag$. Now Bob performs a local unitary operation $\hat{U}_m$ on the projected mixed state $\hat{\varrho}_B$. The outcome is $\hat{\rho}_B=\hat{U}_m\hat{\varrho}_B\hat{U}_m^\dag$, and the corresponding densty matrix is $\rho_B=U_mL_m\rho_C L_m^\dag U_m^\dag$. Rewrite
\begin{equation}
\label{mixedstateB2}
\hat{\rho}_B = p_m^{-1}\sum_{jj^\prime} (U_mL_m\rho_C L_m^\dag U_m^\dag)_{jj^\prime} |j\rangle_B\langle j^\prime|
\end{equation}
Comparing Eq.(\ref{mixedstateB2}) and Eq.(\ref{mixedstateC}), we find that for $\hat{\rho}_B=\hat{\rho}_C$, i.e., achieving faithful teleportation, the condition is (omitting the index $m$) $p^{-1} UL\rho_C L^\dag U^\dag = (p^{-1/2}UL)\rho_C (p^{-1/2}UL)^\dag = \rho_C$. This condition can be obtained if $p^{-1/2}UL=p^{-1/2}UQ^TD^* = I$, which gives $D_m = p_m^{1/2}(Q^{-1})^\dag U_m^T$. But this is exactly the same condition Eq.(\ref{condition1}) for faithfully teleporting a pure state. This can be explained by the fact that Eq.(\ref{mixedstateC2}) can actually represent either a mixed state when $\rho_C^2 \ne \rho_C$, or a pure state when $\rho_C^2 = \rho_C$.

To reformulate the teleportation conditions using density matrix, let's rewrite the reduced density matrix for particle A of the entangled pair AB as $\rho_q=\rho_A^{\cal F}=QQ^{\dag}$ to reflect that it describes the characteristic of the quantum channel. Similarly, the reduced density matrix for particle A of the measurement basis  $|\psi^m_{AC}\rangle$ is $\rho_m=D_mD_m^{\dag}$. Substitute $D_m$ from Eq. (\ref{condition1}), we get $\rho_m=p(Q^{-1})^{\dag}(Q^{-1})=p(QQ^{\dag})^{-1} =p\rho_q^{-1}$, this gives
\begin{equation}
\label{entmatching}
\rho_m\rho_q=pI
\end{equation}
which is the condition of entanglement matching between the quantum channel and the measurement basis. The probability in Eq.(\ref{probability}) is rewritten as by noting that $Tr((Q^{\dag}Q)^{-1}) = Tr((QQ^{\dag})^{-1}) = Tr(\rho_q^{-1})$:
\begin{equation}
\label{probability11}
p = (Tr(\rho_q^{-1}))^{-1}
\end{equation}
Per Lemma 1, $p$ is an invariant under unitary transformation, therefore Eq. (\ref{probability11}) holds true for teleportation with any arbitrary entangle quantum channel. Lastly, taking complext conjugate of Eq. (\ref{condition22}) we have
\begin{equation}
\label{condition23}
Tr(U_n\rho_q^{-1}U_m^\dag) = p^{-1}\delta_{nm}
\end{equation}
We can validate these results by considering a maximum entangled quantum channel, $\rho_q = N^{-1}I$, $\rho_q^{-1} = NI$, we have $p=N^{-2}$, $\rho_m=N^{-1}I$, and Eq.(\ref{condition23}) is simplified to $Tr(U_nU_m^\dag)=\delta_{nm}$. In this case, $\eta=N^2$, and $p_{max} = \eta p = 1$.

The following theorem summarizes the main conclusions of this paper:
\begin{theorem}
The condition for faithful teleportation of a pure state or a mixed state can be described by
\begin{equation}
\label{conclusion}
\rho_m\rho_q=pI, \quad \textrm{and} \quad  p = (Tr(\rho_q^{-1}))^{-1}
\end{equation}
\label{theoryrho}
where $\rho_q$ is reduced density matrix of the quantum channel, $\rho_m$ is the reduced density matrix of the local measurement basis, I is the unit matrix, and p is the probability.
\end{theorem}

\section{\label{NEqual2} Partially Entangled Quantum Channel ($N=2$)}

For any set of orthogonal measurement basis, how many of them lead to faithful teleportation? For partially entangled quantum channel, there is no straightforward solution for Eq.(\ref{condition23}). We will need to rely on the actual parametrization of the unitary matrix and hence will be restrictive to the case of $N=2$. It has been pointed out that when partially entangled bipartite is used as shared resource, there can be two such measurements~\cite{Agrawal02, Pati08}. Reference~\cite{Pati08} also provided a sketch of explanation. However, there is no rigorous proof yet on the fact that a third such orthogonal measurement cannot be found. In this section, a general proof is provided.

\begin{theorem} 
\label{theoryNeq2}
For teleportation using partially entangled quantum channel described by $|\Psi\rangle_{AB}=q_0|0\rangle_A|0\rangle_B + q_1|1\rangle_A|1\rangle_B$ ($|q_{0}| \neq |q_{1}|$), $\eta$ is no greater than two.
\end{theorem}
Finding $\eta$ is equivalent to finding the number of unitary matrices that satisfy Eq. (\ref{condition23}). Since there is no restriction of the unitary operation Bob can perform, we should assume the most generic two dimensional unitary matrix for Eq.(\ref{condition1}),
\begin{equation}
\label{unitary}
U_m = e^{i\phi_m}\begin{pmatrix} \cos\theta_m e^{i\alpha_m} & \sin\theta_m e^{i\beta_m} \\ -\sin\theta_m e^{-i\beta_m} & \cos\theta_m e^{-i\alpha_m} \end{pmatrix}
\end{equation}
where $\phi, \alpha, \beta, \theta \in [0, 2\pi]$ are independent variables. Since $\rho_A^{-1}=\begin{pmatrix}1/|q_0|^2 & 0 \\ 0 & 1/|q_1|^2 \end{pmatrix}$, substitute $U_m, U_n$, and $\rho_A^{-1}$ into Eq. (\ref{condition23}), and with simple algebra, one obtains the following condition
\begin{equation}
\label{condition3}
\begin{split}
Tr & (U_n\rho_A^{-1}U_m^\dag) =|q_1|^{-2}e^{i\phi_{nm}} (g\sin\theta_m\sin\theta_ne^{i\beta_{nm}} \\
& + \sin\theta_m\sin\theta_n e^{-i\beta_{nm}}
 +g\cos\theta_m\cos\theta_ne^{i\alpha_{nm}} \\
& + \cos\theta_m\cos\theta_n e^{-i\alpha_{nm}}) = p^{-1}\delta_{nm}
\end{split}
\end{equation}
where $\phi_{nm}=\phi_n-\phi_m, \alpha_{nm}=\alpha_n-\alpha_m,  \beta_{nm}=\beta_n-\beta_m, g=(|q_{1}|/|q_{0}|)^2$. Obviously when $g=1$, the particles AB are in maximum entangled state. Assuming $\cos\theta_m\cos\theta_n \neq 0$, we define $t_{nm}=(\sin\theta_m\sin\theta_n)/(\cos\theta_m\cos\theta_n)$, Eq. (\ref{condition3}) becomes
\begin{equation}
\label{condition4}
\begin{split}
&Tr(U_n\rho_A^{-1}U_m^\dag)=|q_1|^{-2}e^{i\phi_{nm}}\cos\theta_m\cos\theta_nf_{nm} \\
& f_{nm}=t_{nm}(ge^{i\beta_{nm}}+e^{-i\beta_{nm}})+(ge^{i\alpha_{nm}}+e^{-i\alpha_{nm}})
\end{split}
\end{equation}
When $f_{nm}=0$, $U_n$ and $U_m$ satisfy the orthogonal condition in Eq. (\ref{condition23}), both the real part and imaginary part of $f_{nm}$ equal to zero, therefore
\begin{equation}
\label{condition5}
\begin{split}
& (1+g)(\cos\alpha_{nm}+t_{nm}\cos\beta_{nm})=0 \\
& (1-g)(\sin\alpha_{nm}+t_{nm}\sin\beta_{nm})=0
\end{split}
\end{equation}
Since we assume particles AB are in partially entangle state, $g\neq 1$, and by definition $g>0$, therefore Eq. (\ref{condition5}) is simplified as
\begin{equation}
\begin{split}
\label{condition6}
& \cos\alpha_{nm}+t_{nm}\cos\beta_{nm}=0 \\
& \sin\alpha_{nm}+t_{nm}\sin\beta_{nm}=0
\end{split}
\end{equation}
There are two possible solutions for Eq. (\ref{condition6}), $t_{nm}=1$, $\alpha_{nm}=\beta_{nm}+\pi$, or $t_{nm}=-1$, $\alpha_{nm}=\beta_{nm}$. Note that $t_{nm}=1$ implies $\theta_m+\theta_n=\pi/2$, and $t_{nm}=-1$ implies $\theta_m-\theta_n=\pi/2$. In Eq. (\ref{condition3}), if $\cos\theta_m\cos\theta_n=0$, because of the following inequality
\begin{equation}
\label{inequality}
ge^{i\beta}+e^{-i\beta} \neq 0 \quad \textrm{for any $\beta$, given $g\ne1$}
\end{equation}
we have $\sin\theta_m\sin\theta_n=0$. Therefore there are two more possible solutions, $\cos\theta_m=0, \sin\theta_n=0$, or $\sin\theta_m=0, \cos\theta_n=0$. In summary, there are four possible solutions for Eq. (\ref{condition23}) when $N=2$,
\begin{enumerate}
   \item $\cos\theta_m=0, \sin\theta_n=0$
   \item $\sin\theta_m=0, \cos\theta_n=0$
   \item $\theta_m+\theta_n=\pi/2$, $\alpha_{nm}=\beta_{nm}+\pi$
   \item $\theta_m-\theta_n=\pi/2$, $\alpha_{nm}=\beta_{nm}$.
\end{enumerate}
Consider case 1. Let's choose two unitary matrix $U_1$ and $U_2$ such that $\cos\theta_1=0, \sin\theta_2=0$, then $U_1$ and $U_2$ satisfy Eq. (\ref{condition23}). Now let's also choose a third unitary matrix $U_3$ such that $\sin\theta_3=0$, $U_3$ and $U_1$ also satisfy Eq. (\ref{condition23}). However, given the inequality in (\ref{inequality}) and the fact $\cos\theta_3\cos\theta_2=\pm1$, we have
\begin{equation*}
Tr(U_3\rho_A^{-1}U_2^\dag)=\pm|q_{1}|^{-2}e^{i\phi_{32}}(ge^{i\alpha_{32}}+e^{-i\alpha_{32}}) \neq 0
\end{equation*}
Therefore $U_3$ and $U_2$ do not satisfy Eq. (\ref{condition23}). This means once the two unitary matrices are chosen, it is not possible to find a third one that satisfies Eq. (\ref{condition1}) and (\ref{condition2}). The same argument goes to case 2.

Now consider case 3, we choose $U_1$ and $U_2$ such that $\theta_1+\theta_2=\pi/2$, $\alpha_{21}=\beta_{21}+\pi$. $U_1$ and $U_2$ satisfy Eq. (\ref{condition23}). We also want to choose $U_3$ such that $U_3$ and $U_1$ satisfy Eq. (\ref{condition23}), and there are two sub cases here. a.) $\theta_1+\theta_3=\pi/2$, $\alpha_{31}=\beta_{31}+\pi$; b.) $\theta_3-\theta_1=\pi/2$, $\alpha_{31}=\beta_{31}$. For case 3.a, one has $\theta_3=\theta_2$ therefore $t_{32}=(\tan\theta_3)^2$, and since $ \alpha_{31}-\alpha_{21}=\beta_{31} +\pi -\beta_{21}-\pi$, we have $\alpha_{32}=\beta_{32}$. Eq. (\ref{condition4}) becomes
\begin{equation*}
f_{32}=((\tan\theta_3)^2+1)(ge^{i\beta_{32}}+e^{-i\beta_{32}}) \neq 0
\end{equation*}
For case 3.b, one has $\theta_3=\pi-\theta_2$ therefore $t_{32}=-(\tan\theta_3)^2$. Similarly $ \alpha_{31}-\alpha_{21}=\beta_{31}-\beta_{21}-\pi$, we have $\alpha_{32}=\beta_{32} - \pi$. Eq. (\ref{condition4}) becomes
\begin{equation*}
f_{32}=-((\tan\theta_3)^2+1)(ge^{i\beta_{32}}+e^{-i\beta_{32}}) \neq 0
\end{equation*}
Therefore for case 3, once two unitary matrix are chosen, it is not possible to find a third matrix that satisfies (\ref{condition23}). The same argument goes to case 4. This concludes that in any case, $\eta \le 2$.

From the above proving process, it is clear that the condition $g\neq 1$, which means the quantum channel is not in maximum entangled state, plays a crucial role to restrict the possibility of finding a third orthogonal measurement basis. 
\begin{corollary}
Using partially entangled quantum channel of $|\Psi\rangle_{AB}=q_{0}|0\rangle_A|0\rangle_B + q_{1}|1\rangle_A|1\rangle_B$ where $|q_{0}| \neq |q_{1}|$, the maximum probability of faithful teleportation is given by $p_{max}=2|q_{0}q_{1}|^2$.
\end{corollary}
This is easy to derived. Since $Tr(\rho_A^{-1})=1/|q_0|^2 + 1/|q_1|^2 = 1/|q_0q_1|^2 $ , the probability from each successful teleportation is $p=|q_0q_1|^2$. It is the same for each orthogonal measurement basis when teleportation is successful. Since $\eta \le 2$ according to Theorem \ref{theoryNeq2},  $p_{max}=2p=2|q_0q_1|^2$.
\begin{corollary}
For teleportation using arbitrary partially entangled quantum channel Eq.(\ref{partent}), $\eta \le 2$.
\end{corollary}
Corollary 3.2 is an obvious conclusion by combining Theorem \ref{theoryeta} and \ref{theoryNeq2}. To calculate the total probability of successful teleportaiton for an arbitrary partially entangled quantum channel described by Eq.(\ref{partent}), one first converts Eq. (\ref{partent}) to Eq.(\ref{entG}), then applies Corollary 3.1.  

\section{\label{examples}Operational Examples}
In this section, we show how the theory developed in previous section can be applied to find out the matching measurement basis for any given partially entangled quantum channel, and how the faithful teleportation probability can be computed, in a simple procedure.

First let's consider the quantum channel between Alice and Bob takes the following diagonal form:
\begin{equation}
\label{example1}
|\Psi\rangle_{AB}=\cos\theta|0\rangle_A|0\rangle_B + \sin\theta|1\rangle_A|1\rangle_B
\end{equation}
where $\theta\in[0,\pi]$, and $\theta\ne\pi/4$ or $3\pi/4$ hence not in maximum entanglement. We can only find no more than two orthogonal von-Neumann measurement basis. Let's choose $U_1= I$ and $U_2= \begin{pmatrix} 0&1 \\ -1 & 0 \end{pmatrix}$, The resulting $D_1=\begin{pmatrix} \sin\theta &0 \\ 0 & \cos\theta \end{pmatrix}$ and $D_2=\begin{pmatrix}  0 & -\sin\theta \\ \cos\theta & 0 \end{pmatrix}$. The other two matrices are chosen to satisfy the orthogonal condition but not meet the faithful teleportation condition. $D_3=\begin{pmatrix} \cos\theta & 0 \\ 0 & -\sin\theta \\ \end{pmatrix}$, and $D_4=\begin{pmatrix} 0 & \cos\theta \\ \sin\theta & 0 \end{pmatrix}$. The four orthogonal measurement basis Alice can performed are:
\begin{equation}
  \label{mbasis2}
  \left\{
  \begin{split}
& |\psi^1_{AC}\rangle= \sin\theta|0\rangle_A|0\rangle_C + \cos\theta|1\rangle_A|1\rangle_C \\
& |\psi^2_{AC}\rangle= -\sin\theta|0\rangle_A|1\rangle_C + \cos\theta|1\rangle_A|0\rangle_C \\
& |\psi^3_{AC}\rangle= \cos\theta|0\rangle_A|1\rangle_C + \sin\theta|1\rangle_A|0\rangle_C \\
& |\psi^4_{AC}\rangle = \cos\theta|0\rangle_A|0\rangle_C - \sin\theta|1\rangle_A|1\rangle_C
\end{split}
\right.
\end{equation}
The first two measurement sets yield faithful teleportation, while the last two don't. The total probability of faithful teleportation with this quantum channel is:
\begin{equation}
\label{probability2}
p_{max}=2(\cos\theta\sin\theta)^2=\frac{1}{2}\sin^2(2\theta)
\end{equation}

If the coefficients of the arbitrary partially entangled state in Eq.(\ref{partent}) are real numbers and in the case of $N=2$, we can define the $2\times2$ unitary matrix in the following format
\begin{equation}
\label{unitary2}
U_A = \begin{pmatrix} \cos\theta_A & -\sin\theta_A \\ \sin\theta_A & \cos\theta_A \end{pmatrix}, U_B = \begin{pmatrix} \cos\theta_B  & -\sin\theta_B \\ \sin\theta_B & \cos\theta_B \end{pmatrix}
\end{equation}
where $\theta_A$ and $\theta_B$ are independent variables. It is easy to prove that there always exist solutions for $\theta_A$ and $\theta_B$ such that $U_A$ and $U_B$ can transform the arbitrary partially entangled state to the Schmidt decomposition form Eq.(\ref{entG}). In the second example, we consider a quantum channel that is described by
\begin{equation}
\label{example2}
\begin{split}
|\Psi\rangle_{AB}^{\cal F} = & -0.1|0\rangle_A|0\rangle_B - 0.7 |0\rangle_A|1\rangle_B \\
& +0.7|1\rangle_A|0\rangle_B + 0.1|1\rangle_A|1\rangle_B.
\end{split}
\end{equation}
We define two local unitary transformations described in Eq.(\ref{unitary2}) such that $\theta_A=\pi/4$ and $\theta_B=-\pi/4$. This transforms Eq. (\ref{example2}) to
\begin{equation}
  \label{example2G}
|\Psi\rangle^{\cal G}_{AB}=0.6|0\rangle_A^\prime|0\rangle_B^\prime+0.8|1\rangle_A^\prime|1\rangle_B^\prime.
\end{equation}
Therefore, $p_{max}=2(0.6)^2(0.8)^2=0.4608$. The four orthogonal measurement basis that Alice can perform on particles AC, after converting back to the ${\cal F}$ picture, are
\begin{equation*}
  \label{mbasis3}
  \left\{
  \begin{split}
& |\psi^1_{AC}\rangle= 0.1|0\rangle|0\rangle - 0.7 |0\rangle|1\rangle +0.7|1\rangle|0\rangle - 0.1|1\rangle|1\rangle \\
& |\psi^2_{AC}\rangle= -0.7|0\rangle|0\rangle - 0.1 |0\rangle|1\rangle -0.1|1\rangle|0\rangle - 0.7|1\rangle|1\rangle \\
& |\psi^3_{AC}\rangle= -0.1|0\rangle|0\rangle + 0.7 |0\rangle|1\rangle +0.7|1\rangle|0\rangle - 0.1|1\rangle|1\rangle \\
& |\psi^4_{AC}\rangle= 0.7|0\rangle|0\rangle +0.1 |0\rangle|1\rangle -0.1|1\rangle|0\rangle - 0.7|1\rangle|1\rangle \\
\end{split}
\right.
\end{equation*}
where the first two measurement sets $|\psi^1_{AC}\rangle$ and $|\psi^2_{AC}\rangle$ achieve faithful teleportation.

\section{\label{discussion}Discussion}

Consider the total probability of faithful teleportation in Eq. (\ref{probability2}). When $\theta\to\pi/4$ or $3\pi/4$, $p_{max}\to0.5$. However, when $\theta=\pi/4$ or $3\pi/4$, $p_{max}=1$ instead of $0.5$. This means when the pair becomes maximum entangled, the maximum probability of faithful teleportation jumps to 1 instead of 0.5. To explain this, we can separate the factors that determine the faithful teleportaion into two. One is the tuning of the LOCC to the quantum channel, i.e., the entanglement matching~\cite{Li00}. The other is the degree of entanglement of the quantum channel itself. When the quantum channel is in maximum entangled state, the measurement basis (corresponding to the LOCC) can be chosen to perfectly match the entanglement of the quantum channel, thus achieves unity probability. However, when the quantum channel is partially entangled, the LOCC can only be matched up to two orthogonal measurement basis, i.e., there is only one bit of useful classical information can be sent to Bob. Even with the entanglement matching, the quantum channel itself is partially entangled, the maximum probability is less than 0.5. The less of degree of entanglement, the less of the probability of faithful teleportation. 

Experimentally it should be straightforward to confirm the probability in Eq.(\ref{probability2}). For example, using the experiment setup in Ref.~\cite{Kim02}, for a shared partially entangled pair described in Eq.(\ref{example1}), one can just tilt the four detectors appropriately according to Eq.(\ref{mbasis2}), and measure the successful teleportaiton rate from two of the detectors. This should confirm the total probability given by Eq.(\ref{probability2}).  

In summary, we show the probability of faithful teleportation equals to the inverse of the trace of the inverse of reduced density matrix of the quantum channel. When teleportation is successful, the reduced density matrix of the quantum channel and the reduced density matrix of the measurement basis are inverse each other with a factor equal to the probability. These are general results applicable to teleporting either a pure state or a mixed state. We also provide a rigorous proof that using arbitrary partially entangled four dimensional quantum channel ($N=2$ and $N^2=4$), for any given orthogonal measurement basis, only up to two of them can faithfully teleport a qubit. The faithful teleportation conditions are invariant under unitary transformation. The significance of this result may not be theoretical but rather operational. This can be seen from the example in section \ref{examples} which shows a simple procedure to find the faithful teleportaiton probability and the matching measurement basis, for any partially entangled quantum channel.

\section*{Acknowledgement}
The author gratefully thanks Qualcomm and UCSD Library for granting access to the research journals and database.

\end{document}